\documentclass[%
 reprint,
 amsmath,amssymb,
 aps,
]{revtex4-2}

\usepackage{graphicx}
\usepackage{dcolumn}
\usepackage{bm}
\usepackage{adjustbox}
\usepackage{multirow}
\setcitestyle{numbers,square}

\usepackage{xcolor}

\begin{document}

\preprint{APS/123-QED}

\title{A subset selection based approach to finding important \\structure of complex networks}

\author{Richa Tripathi}
\email{richa.tripathi@iitgn.ac.in}
 \affiliation{Indian Institute of Technology Gandhinagar, Gujarat, India}
\author{Amit Reza}%
\affiliation{%
 Indian Institute of Technology Gandhinagar, Gujarat, India}%

\date{\today}

\begin{abstract}
Most of the real world networks such as the internet network, collaboration networks, brain networks, citation networks, powerline and airline networks are very large and to study their structure, and dynamics one often requires working with large connectivity (adjacency) matrices. However, it is almost always true that a few or sometimes most of the nodes and their connections are not very crucial for network functioning or that the network is robust to a failure of certain nodes and their connections to the rest of the network. In the present work, we aim to extract the size reduced representation of complex networks such that new representation has the most relevant network nodes and connections and retains its spectral properties. To achieve this, we use the Subset Selection (SS) procedure. The SS method, in general, is used to retrieve maximum information from a matrix in terms of its most informative columns. The retrieved matrix, typically known as subset has columns of an original matrix that have the least linear dependency. We present the application of SS procedure to many adjacency matrices of real-world networks and model network types to extract their subset. The subset owing to its small size can play a crucial role in analyzing spectral properties of large complex networks where space and time complexity of analyzing full adjacency matrices are too expensive. The adjacency matrix constructed from the obtained subset has a smaller size and represents the most important network structure. We observed that the subset network which is almost half the size of the original network has better information flow efficiency than the original network. Also, we found that the contribution to the Inverse Participation ratio of the network comes almost entirely from nodes that are there in the subset. This implies that the SS procedure can extract the top most influential nodes without the need for analyzing the full adjacency matrix.

\end{abstract}

\maketitle


\section{Introduction}
A wide varieties of real world systems have been studied using the network framework - communication systems networks such as the Internet\cite{albert1999internet}, transport and power grid networks\cite{pagani2013power} - social networks such as Twitter \cite{ediger2010massive}, Facebook\cite{catanese2011crawling} and collaboration networks\cite{newman2001structure} - networks at the cellular scale such as metabolic networks\cite{ravasz2002hierarchical} and protein interaction networks\cite{liu2009complex}. Such a framework has helped us infer network function from the structure and vice-versa \cite{newman2003structure}. For example - finding of the most influential disease spreader in epidemic spreading \cite{kitsak2010identification}, finding the most versatile author in collaboration network\cite{de2015ranking} or finding the most important centre for information processing in brain\cite{van2013network}. One of the prior challenges to analysis and visualization of these networks is their enormous size \cite{albert2002r}. Any algorithm that uses network topology as input, runs on time that grows polynomially with network size\cite{clauset2004finding}, \cite{newman2004finding}. One possible solution is to coarse grain the network such that its size is reduced and important network information in terms of significant nodes and edges is retrieved\cite{gfeller2007spectral}, \cite{kim2011reduction}, \cite{arenas2007size}.\\

Recently, there have efforts to infer structural reducibility of complex networks such that their functioning remains intact\cite{kim2011reduction}. This possibility breeds on the existence of unimportant network structure or the presence of redundancy in the network data. The robustness of most real-world networks such as the Internet and power grid networks, despite several targeted/un-targeted attacks, the presence of malfunctioning nodes and local failures are attributed to the presence of redundant network wiring\cite{albert2000error}. There has been a recent study on structure reducibility of multi-layer networks \cite{de2015structural} where it is shown that up to $75$ percent of the network is redundant for representing accurate multi-layer network structure.\\

In the present work, we show the applicability of the subset selection algorithm for structure reduction or summarization of complex networks. This can be mapped to a problem of finding a summary of a document set as stated in\cite{chen2009summarizing}: Imagine a network with nodes representing the sentences of the document and edges their weighted similarities. The summary of the document has a smaller number of sentences than the original document, and the sentences therein are each rich in information and maximally diverse from each other. Hence, the summary network of the document has the most important nodes and their connectivities to other important nodes. The subset selection is analogous to the concept of feature selection \cite{john1994irrelevant}, \cite{boutsidis2009improved} in large data sets which uses techniques such as Principal Component Analysis (PCA), where relevant features are selected capturing the rich and the most diverse features out of all. The advantage of feature selection is in the convenience of handling a small subset of the full data with most of the information as the original data set and with non-redundant structure. Such subset selection procedures have been used extensively in training the feed-forward neural networks \cite{kanjilal1995application} for modeling time series from dynamical systems. It has found many applications in solving rank-deficient least square problems \cite{golub2012matrix}, in genetics \cite{butler2005strategies}, in wireless communication \cite{wilzeck2008antenna} and other information retrieval problems \cite{boutsidis2009improved}, \cite{chandrasekaran1994rank}.\\

The subset selection procedure as carried out in the present work captures the $q$(say) important network nodes and their connections. The quantity $q$ is the pre-specified number of columns that are extracted from the original network adjacency matrix into the subset. These columns would be the most linearly independent column set from the original matrix and hence represent the unique and most diverse set. This column set can in principle (provided $q$ appropriately chosen) be used to represent the whole matrix, as the left out columns (redundant set) are mostly linearly dependent. Hence, the determination of $q$ is an essential aspect of the subset selection procedure. Also, it is known that the matrix rank gives the number of the most linearly independent set of vectors. Since we are looking for $q$ representative columns of the whole matrix, the value $q$ is straightforwardly the numerical rank of the matrix. The subset selection is most significant for rank deficient matrices; the rank deficiency is implying the linear dependency of columns. As any rank deficient matrix has redundant structure, it can be represented by a smaller number of columns(the subset captures that). Hence, the adjacency matrices which have linearly dependent columns or correspondingly a network that has a redundant network structure are the best targets of the subset selection procedure. The power of the SS procedure will tremendously increase in case of sparse networks. Moreover, if there are disconnected nodes, then the adjacency matrix contains a large number of zero-columns which increase the amount of redundancy in the adjacency matrix. Therefore $q$ will be very small in comparison to the total number of nodes and the obtained reduced adjacency matrix will be a small matrix. There is a clear trade-off between the sparsity in the network structure and $q$. Less sparsity implies the moderately high value of $q$, and as a result, the reduced adjacency will also become large. Hence, SS procedure recommended for applying to sparse adjacency matrices.

On a different note, if one requires to find important network structure comprising of $q$ nodes, where $q$ is arbitrary or user-defined, the subset selection procedure can be employed to find it. We show in the paper the effect of choosing different $q$'s in terms of information retrieval from the original network. The preserved matrix norm in the subset and the overlap of Principal Singular Vector of the original matrix and the subset quantify this information retrieval.  \\

The classic Subset Selection ($SS$) procedure as in \cite{golub2012matrix} uses QR-column pivoting factorization \cite{higham2000qr} on  a matrix of the first few right singular vectors \cite{golub1971singular} (say $q$) corresponding to the top $q$ singular values to obtain a permutation vector $P$ \cite{chan1987rank}. The $P$ vector is then employed to rank the original matrix columns in order of their importance, and the selected subset comprises of first $q$ columns of the ordered matrix. We apply this procedure on the test networks generated artificially and on a few real-world networks. We also compare the network properties of the original network and the subset network and find that the subset network is more efficient in information flow. The efficiency is quantified in terms of the network metrics \cite{albert2002statistical}.\\

\section{Subset Selection Procedure}
The subset selection procedure is generally applied to identify the most important column vectors of a data matrix $\mathcal{A}$ in such a way that the obtained subset retrieves the maximum amount of information from the matrix. Or, one identifies those columns of $\mathcal{A}$ such that the energy of $\mathcal{A}$ is optimally preserved. The matrix, $\mathcal{A}$ can be thought of as a collection of two blocks $[\mathcal{A}_1, \mathcal{A}_2]$ after applying subset selection procedure onto $\mathcal{A}$. Where $\mathcal{A}_1$ contains $q$  most linearly independent columns which can span the entire column space of $\mathcal{A}$ and the remaining block $\mathcal{A}_2$ contains the redundant columns which are well represented by the linear combination of $\mathcal{A}_1$ s.t. $\textrm{min}_{x} \|\mathcal{A}_1\, x - \mathcal{A}_2 \|_2$ is very small.
The concept of rank-deficiency of $\mathcal{A}$ plays a crucial role in deciding the number of independent columns present in $\mathcal{A}_1$. More the rank-deficiency of $\mathcal{A}$ less is the number of columns in $\mathcal{A}_1$ and vice-versa.
Hence, the subset selection procedure is all about to finding out the non-redundant (representative) block $\mathcal{A}_1$ and redundant block $\mathcal{A}_2$ respectively. The permutation matrix $P$ can synchronize the representative columns of $\mathcal{A}_1$ together. Therefore, using $P$ it is possible to obtained the left- side block and right-side block representation of $\mathcal{A}$ as follows.
\begin{equation}
\label{Perm}
\mathcal{A}\, P = [\mathcal{A}_1, \mathcal{A}_2]
\end{equation}
Apparently, the whole subset selection procedure transforms into a problem of finding the optimal permutation matrix $P$ in such a way that it follows following desirable constraints.
\begin{enumerate}
    \item The number of linearly independent columns ($q$)of $\mathcal{A}_1$ should represent the optimal rank of $\mathcal{A}$, i.e., $q$ should tend towards the numerical rank.
    \item The residual difference between the norm of the linear combination of $\mathcal{A}_1$ with $\mathcal{A}_2$ should be minimal.
\end{enumerate}
The successful incorporation of these two constraints helps to obtain an optimal $\mathcal{A}_1$ which contains the maximum information of $\mathcal{A}$.
The constraints can be satisfied by studying the singular values spectra of $\mathcal{A}$ which provides an optimal way for finding the number of the linearly independent set of columns. Hence, the optimal numerical rank can be decided based on the top-$q$ non-singular values. 
Therefore the first pre-requisite of sub-set selection algorithm is to apply singular value decomposition (SVD) on $\mathcal{A}$ to obtain the top-$q$ singular values and corresponding singular vectors.
The SVD of a typical $\mathcal{A}_{m \times n}$ matrix results in three matrices ( $\mathcal{U}, \Sigma,  \mathcal{V}$) such that
\begin{equation}
\label{svd}
\mathcal{A} = \mathcal{U} \Sigma \mathcal{V}^T.
\end{equation}

The matrix $\mathcal{U}_{m \times r}$ represents eigenvectors of the left subspace of $\mathcal{A}$; the matrix $\Sigma_{r \times r}$ is a diagonal matrix with $r$ ($r = rank(\mathcal{A}) = min(m,n)$) positive non-zero entries (known as singular values) arranged in descending order of magnitude i.e ($\sigma_1 > \sigma_2 > \sigma_3 > ...> \sigma_r$) and the matrix $\mathcal{V}_{r \times n}$ represents eigenvectors of the right subspace of $\mathcal{A}$. The energy of the matrix is represented by its Frobenius norm which is defined as the square root of squared sum of all the $r$ singular values. However, there can be matrices for which not all $r$ singular values are dominant (or have zero or negligible magnitude). This means that the Forbenius norm is almost fully expressed by $q$ out of $r$ singular values i.e $\sum_{i = 1}^{q} \sigma_i^2  \simeq \sum_{i = 1}^{r} \sigma_i^2 = \| \mathcal{A} \| _F^2$,
where $\| \mathcal{A} \|_F^2$ is the Frobenius norm of $\mathcal{A}$.\\

From eq.\ref{Perm} $\&$ \ref{svd}, it is clear that permuting columns of $\mathcal{A}$ can be obtained by
permuting columns of the right singular vector matrix $\mathcal{V}$ after truncation upto $q$ columns .
This means that selecting important columns from $\mathcal{A}$ is equivalent to selecting the corresponding columns in $\mathcal{V}^T$. Therefore, the second step of the SS procedure is to obtain the permutation matrix based on truncated $\mathcal{V}^T$. Let $\bar{V}$ represent a matrix of first $q$ columns $\mathcal{V}$ and $\bar{V}^T$ is its transpose. The standard SS method uses $QR$ factorization with column pivoting ($QR_{cp}$) on the matrix $\bar{V}^T$ to obtain a permutation vector $P$.\\
This is represented mathematically as follows,

\begin{equation}
\bar{V}^T = QRP^T
\end{equation}

where $Q$ is a matrix of orthogonal vectors, and $R$ is an upper triangular matrix. In a general QR decomposition of a matrix $A_{q \times n}$, if $A$ has $k$ ($\leq n$) linearly independent columns then the first $k$ columns of $Q$ span first $k$ columns of $A$. Any $k^{\text{th}}$ column of $A$ depends only on first $k$ columns of $Q$. Hence $R$ has an upper diagonal structure.
The $QR_{cp}$ of $\bar{V}^T$ (in addition to general QR) finds $P$ such that the diagonal elements of $R$ are non increasing i.e $|r_{11}| \geq |r_{22}|..... \geq |r_{nn}| $. $P$ is then used to order columns of $\mathcal{A}$ such that most linearly independent columns come first and linearly dependent ones come at the last (eq.~\ref{eqn:eqp}).

\begin{equation}
[\mathcal{A}_1, \mathcal{A}_2] \equiv \mathcal{A}P
\label{eqn:eqp}
\end{equation}

where $\mathcal{A}_1$ is a $m \times q$ matrix representing important and reduced structure of the original network matrix, $\mathcal{A}$ and $\mathcal{A}_2$ (of size $m \times n-q$)represents redundant structure. The subset hence obtained is a rectangular matrix of size $m \times q$.\\

\subsubsection*{Obtaining subset of the Network Adjacency Matrix}
In this sub-section, we obtain the reduced adjacency matrix by applying SS procedure on the adjacency matrix $\mathcal{A}$ of complex networks. As a first step of the SS procedure, SVD of $\mathcal{A}$ is performed (ref.eq.\ref{svd}).
However, as the adjacency matrices are square matrices, therefore one can alternatively compute the Eigenvalue Decomposition (EVD) of $\mathcal{A}$ i.e $\mathcal{A}_{m \times m} =  X \Lambda X^T$, where $\Lambda$ is the diagonal square matrix with eigenvalues of $\mathcal{A}$ along the diagonal and $X$ is a matrix with orthogonal eigenvectors.
It is trivial to map the factors ($\mathcal{U}$, $\mathcal{U}$ and $\Sigma$) obtained from SVD of $\mathcal{A}$ to EVD. The singular values are the squares of eigenvalues and the matrices $\mathcal{U}$ and $\mathcal{V}$ are both same and equal to $X$.
Although the two decompositions are same for a square matrix, the SVD method orders the singular values according to their magnitudes in the $\Sigma$ matrix and hence the corresponding singular vectors are arranged in order of their importance in  $\mathcal{U}$ and $\mathcal{V}$ matrices. However, this is not the case in EVD. One needs to explicitly order the eigenvalues and hence the corresponding eigenvectors after performing EVD.
Hence, it is clear that one can obtain EVD/ SVD of the adjacency matrix to get the top-$q$ eigenvectors or right singular vectors based on top-$q$ eigenvalues or singular values and proceed further for $QR_{cp}$ to acquire the permutation vector $P$.


\section{Application of SS on Complex Networks}
\begin{figure*}[htbp]
\centering
\begin{tabular}{cc}
 \centering
 \includegraphics[height=5cm]{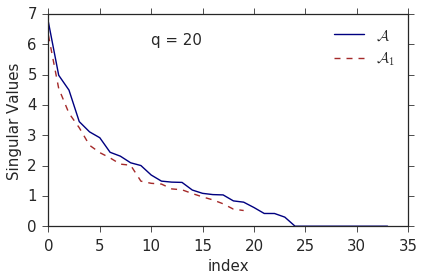}&
 \includegraphics[height=5cm]{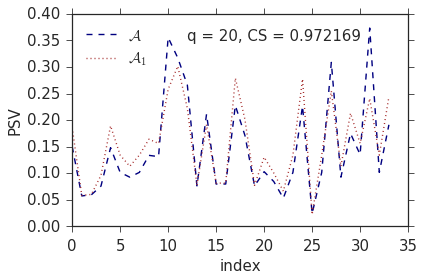}\\
 (a)&(b)
 \end{tabular}
 \caption{Figure showing (a)singular value spectrum and (b) principal singular vectors; of Zachary's Karate club Network (with nodes and edges $N = 34$, $E = 78$) and that of its subset with $q = 20$. }
\label{fig:cs1}
 \end{figure*}
 
\begin{figure*}[htbp]
\centering
\includegraphics[height=10cm, trim={10 60 0 60},clip]{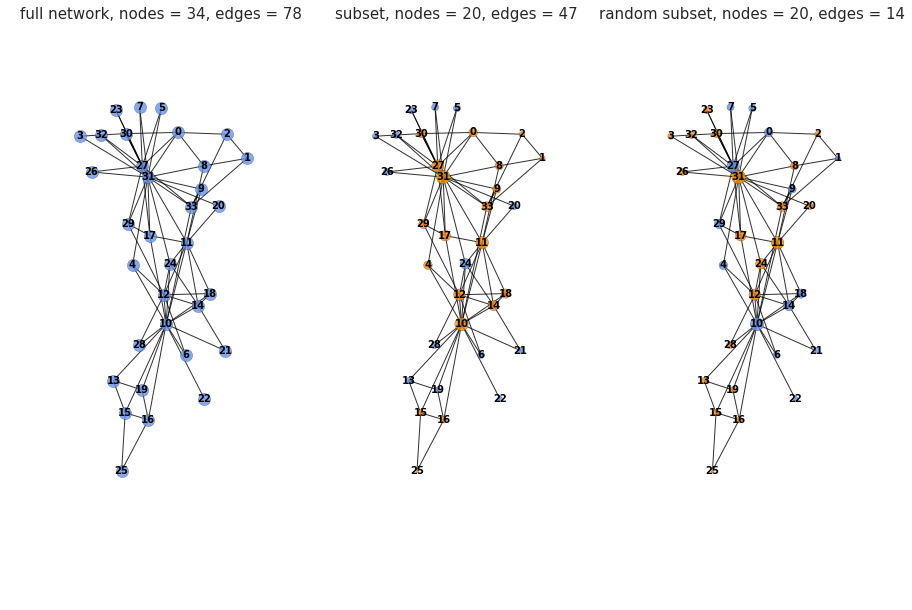}
\caption{The figure showing original network of Zachary's Karate club (first panel), the subset network embedded in the original network (second apnel) and the random subset network embedded in the original network (third panel). The embedded subsets are shown in orange color; the nodes are scaled in size according to values of corresponding components of PSV of $\mathcal{A}$ in panels two and three.}
\label{fig:cs2}
\end{figure*}

An essential aspect to size reduction of complex networks is to check if the eigenvalue spectra are retained in the subset and to quantify any loss in matrix energy in terms of Frobenius norms difference of the $\mathcal{A}$ and the $\mathcal{A}_1$. To this end, we calculated the singular values spectra and left principal singular vector (PSV) of both these matrices in all the examples we present in the following sections.
Ideally, one should compare the PEV for both $\mathcal{A}$ and $\mathcal{A}_1$. But, as $\mathcal{A}_1$ is a non-square matrix therefore, comparison of PEV is not possible and hence the comparison of the PSVs for both the matrices is performed. For a square matrix $\mathcal{A}$,  $\text{PEV}(\mathcal{A}) = \text{PSV}(\mathcal{A})$. Now, if $\text{PSV}(\mathcal{A}) \approx \text{PSV}(\mathcal{A}_1)$, then one can conclude that PEV of $\mathcal{A}$ can be approximated with high accuracy by the PSV of $\mathcal{A}_1$. 
This procedure will help us gauge the relevance of the selected subset in terms of information retrieval from the original network. For example, a network for which the subset selection procedure uses a sufficiently small $q$ (as compared to $m$) to yield a subset that maximally preserves the original matrix energy and has high enough similarity of PSV to the original network can be considered as an apt candidate.\\

On the other hand, one may not require putting above constraint conditions (preservation of full matrix energy in the subset and overlap of PSVs) on the SS procedure at all and compute subset for an arbitrary value of $q$. For such cases, the subset has $q$ most linearly independent columns. Surprisingly, in most of the networks, even when the subset retains only $50 \%$ of the columns, it maximally preserves the norm (see TABLE I.). Also, the PSV of the subset and that of the original network show excellent overlap evaluated in terms of Cosine Similarity of these vectors. The Cosine Similarity (CS) is a measure of relative orientations of the two vectors (refer eq.~\ref{eq:cs}, where $\mathcal{A}^{1}$ and $\mathcal{A}_{1}^{1}$ are PSVs of main adjacency matrix and the subset). Bounded between [0, 1], it is maximum when the two vectors are oriented along the same direction and minimum when they are perpendicular to each other. In our case, the CS is a measure of the extent of information retrieval from the main matrix into the subset.  \\
 
 \begin{equation}
 CS = cos(\theta) = \frac{\mathcal{A}^{1}.\mathcal{A}_{1}^{1}}{\parallel \mathcal{A}^{1} \parallel \parallel \mathcal{A}_{1}^{1} \parallel}
 \label{eq:cs}
\end{equation}

To obtain a network representation of subset or the subset adjacency matrix, we must convert the subset to a square matrix. To this end, we extended the subset selection procedure and reordered rows of the selected subset using $P$(Refer eq.~\ref{eqn:eqB}, $\mathcal{B}$ has top $q$ rows and $\mathcal{B}_r$ has remaining ones). This is justified as the rows, and column vectors in the adjacency matrix are the same, i.e., the adjacency matrix is symmetric (for un-directed networks). Hence the subset adjacency matrix is a square matrix of size $q \times q$, with top q rows and columns retained from the original adjacency matrix. In the following sections of the paper we will refer to original network adjacency as $\mathcal{A}$ (square matrix), the selected subset as $\mathcal{A}_1$ (rectangular matrix) and subset adjacency matrix as $\mathcal{B}$ (square matrix). \\

\begin{equation}
   [\mathcal{B}^T, \mathcal{B}_r^T] \equiv P\mathcal{A}_1
   \label{eqn:eqB}
\end{equation}

\subsection*{Verification of the selected subset}
 To verify whether the subset network is the important/relevant network substructure we calculated the Inverse Participation Ratio (IPR) of the original network. The IPR is defined as $IPR = \sum_{i = 1}^m v_i^4$, where $v_i$ is the eigenvector centrality score of the $i$th node in a network of $m$ nodes. The $IPR \sim \mathcal{O}(1/m)$, implies delocalized state of principal eigenvector (PEV) i.e  $PEV =(1/\sqrt{m}, 1/\sqrt{m}, . . . , 1/\sqrt{m})^T$ and $IPR \sim  \mathcal{O}(1)$ implies complete localization onto a single node. Next, we calculated the contribution to total IPR from those components of PEV, that are indexed by the columns selected in the subset. It is found that the IPR of the original network is almost exactly reproduced by the contributions from the components that are there in the subset. In other words, summing the fourth power of eigenvector centralities of the nodes that are there in the subset ($q$) reproduces IPR, i.e., $\sum_{i \in SS}v_i^4 \approx IPR$, even for subsets that retain only half of the nodes from the network. Since eigenvector centrality values are used to measure node importance, this also implies that network nodes in the main network were not all important or influential or that almost all of the nodes in the selected subset were influential nodes. To summarize, the subset captures the top $q$ influential nodes of the network; influence being measured in terms of their eigenvector centralities \cite{xu2019identification}. Hence, the subset selection procedure can also be posed as an algorithm that finds out top influential nodes in the network.\\

We will illustrate these results through some artificial and real-world network examples below.

\begin{figure*}[htbp]
\centering
\includegraphics[height=10cm, trim={10 60 0 60},clip]{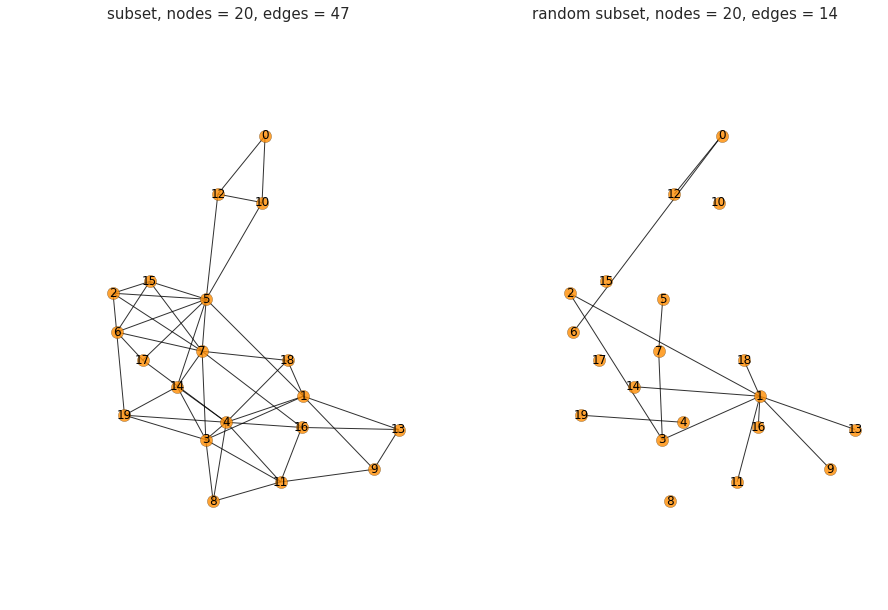}
\caption{The figure showing the actual subset network with 20 nodes and 47 edges (on left) and the random subset network with 20 nodes and 14 edges of Zachary's Karate club network (on right).}
\label{fig:cs3}
\end{figure*}

\begin{figure*}[htbp]
\centering
\begin{tabular}{cc}
 \centering
 \includegraphics[height=5cm, trim={40 60 40 60},clip]{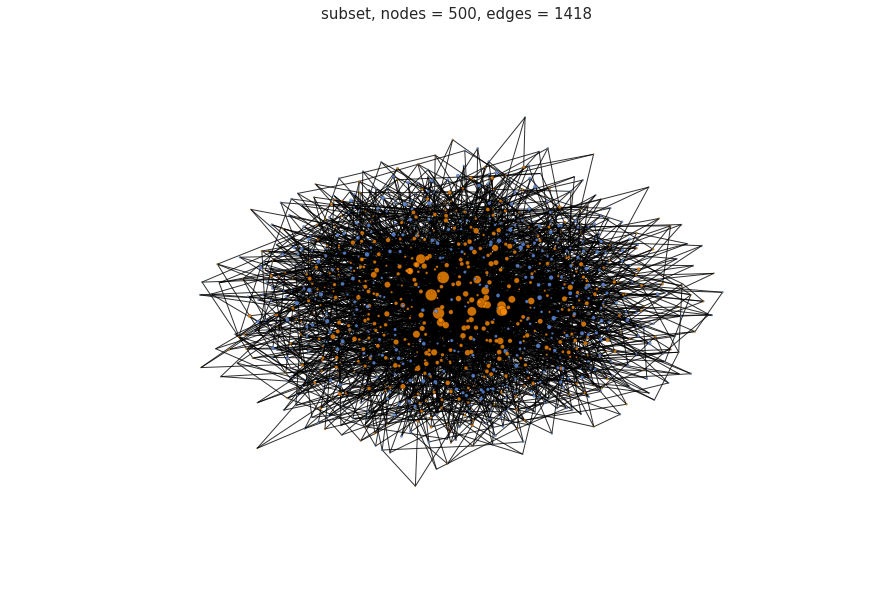}&
 \includegraphics[height=5cm, trim={40 60 40 60},clip]{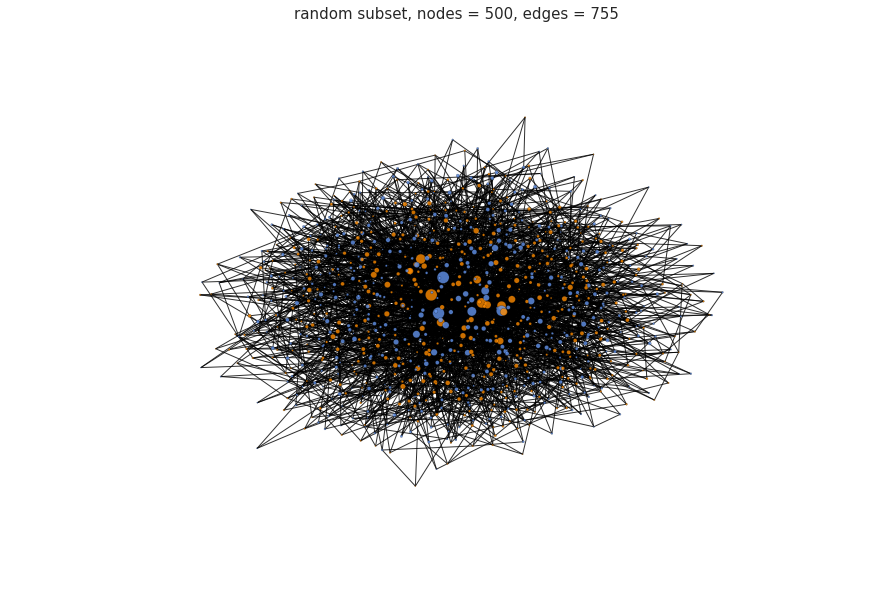}\\
 (a)&(b)
 \end{tabular}
 \caption{The figure showing the original $Barabasi–Albert$(BA) model network with 1000 nodes with embedded subset (first panel) and original network with embedded random subset (second panel). The embedded subsets are shown in orange color. The node sizes in both panels are scaled with corresponding components of PSV of $\mathcal{A}$}
\label{fig:cs6}
 \end{figure*}

\begin{figure*}[htbp]
\centering
\begin{tabular}{cc}
 \centering
 \includegraphics[height=5cm]{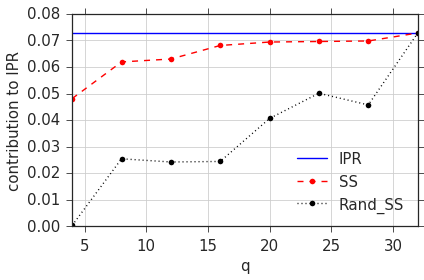}&
 \includegraphics[height=5cm]{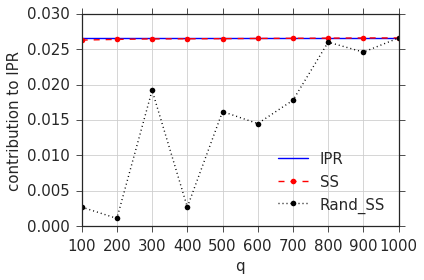}\\
 (a)&(b)
 \end{tabular}
 \caption{Figure showing contribution to total IPR of network, from the subset nodes ($q$) and the random subset nodes for (a)Karate Network and (b) Barabasi- Albert model network. The blue horizontal line is the actual IPR value of the network and red dots and black dots on red and black curves are the contributions of the nodes selected by subset and random subset respectively.}
\label{fig:cs5}
 \end{figure*}
 
 \begin{figure*}[htbp]
\centering
\begin{tabular}{cc}
 \centering
 \includegraphics[height=5cm]{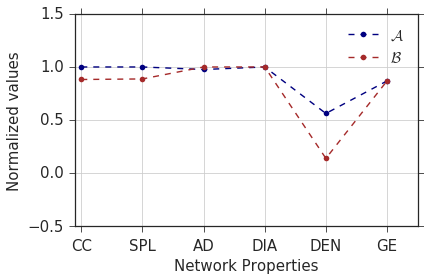}&
 \includegraphics[height=5cm]{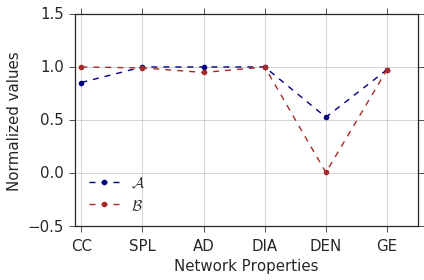}\\
 (a)&(b)
 \end{tabular}
 \caption{The figure showing the normalized network properties of original network and that of subset of (a) the Karate club network, with subset of size $q = 20$, and (b) 1000 node BA network with subset of size $q = 500$}
\label{fig:cs4}
 \end{figure*}


\subsection{Zachary's Karate Club network}
The simplest real-world network example is that of a friendship network of Zachary's Karate club \cite{kunegis2013konect}. The network is unweighted and comprises of $34$ nodes and $78$ edges. We present below the results of subset selection on this network with $q = 20$. The Frobenius norms (Matrix Energies) of the original adjacency matrix and the subsets are $12.49$ and $11.13$ respectively. One can see from singular value spectra (FIG.~\ref{fig:cs1}) of the network that only $25$ out of $34$ singular values are non-zero and contribute to the Frobenius norm. \\

We evaluate the PSVs using the first column of the $\mathcal{U}$ matrix \cite{xu2019identification} of the SVD of $\mathcal{A}$ and $\mathcal{A}_1$. The overlap of their PSVs measured in terms of their Cosine Similarity (CS) (eq.~\ref{eq:cs}) is $0.97$. \\

To evaluate the performance of the SS, we compared the computed subset to a randomly selected subset. A random subset is formed by selecting $q$ columns from $\mathcal{A}$ randomly, without following any order.  While plotting the network, the node sizes were scaled with the corresponding values of components of PSV of $\mathcal{A}$. The values of the PSV components are representative of the node influence and represent eigenvector centralities of nodes in general \cite{ruhnau2000eigenvector}, \cite{bonacich2007some}. We then colored the nodes belonging to the subset networks (true subset and random subset) with a different color. In FIG.~\ref{fig:cs2} the first panel shows the original network (in blue), the second panel shows the subset embedded (in orange) in the original network and third panel shows a random subset (in orange) also embedded in the original network. All the nodes shown in panels two and three are ordered in sizes according to their eigenvector centralities (in the original network corresponding to $\mathcal{A}$). One can see that the subset nodes are the ones that have maximum eigenvector centrality or have maximum influence in the dynamics on networks (Bigger sized nodes being orange). On the other hand, a random subset does not capture all the influential nodes (Bigger Sized nodes not necessarily orange). Hence, the actual subset represents the most important network sub-structure as compared to any random subset.\\

Also, for $\mathcal{B}$ of the actual subset the network is connected and captures $47$ edges as compared to the total of $78$ edges in the network of $\mathcal{A}$. On the other hand, $\mathcal{B}$ calculated from the random subset has only $14$ edges in its network, and the network has many disconnected components (see FIG.~\ref{fig:cs3}). The IPR value of the primary network was $0.073$; the contribution to IPR by nodes marked by subset and the random subset was $0.069$ and $0.042$ respectively.  This implies that the influence localization occurs maximally on the nodes selected by the subset selection procedure. In FIG.\ref{fig:cs5}(a), we show the contributions to IPR of the nodes selected by the subsets of different sizes($q$). We see that beyond $q = 20$, almost full IPR is accounted for by the subset nodes.\\

The network properties of networks corresponding $\mathcal{A}$ and $\mathcal{B}$ and were evaluated and normalized between $0$ and $1$ to compare their relative magnitude (see FIG.~\ref{fig:cs4}(a)). The evaluated network properties\cite{albert2002statistical} were: Clustering Coefficient (CC), Shortest Path Length (SPL), Average Degree (AD), Diameter(DIA), Density(DEN) and Global Efficiency(GE). The decreased SPL and increased GE of the subset network suggest that subset is better in terms of information transmission and can be viewed as effective network sub-structure.\\

In the limit of $q \rightarrow 34$, the subset network will approach the original network, and all the spectral properties of the network adjacency and the structural-functional properties of the network will be accurately reproduced from the subset. The subset selection procedure as demonstrated in this example is carried out on larger and weighted networks, where the subset composed by a $q$ number of essential nodes can effectively find important network nodes and their connections. Also, it is expected that the value of $q$ will be much smaller than the size of the original network.

\begin{figure*}[htbp]
\centering
\begin{tabular}{cc}
 \centering
 \includegraphics[height=5cm]{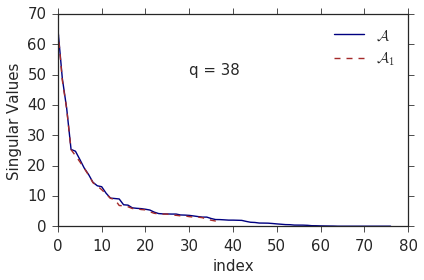}&
 \includegraphics[height=5cm]{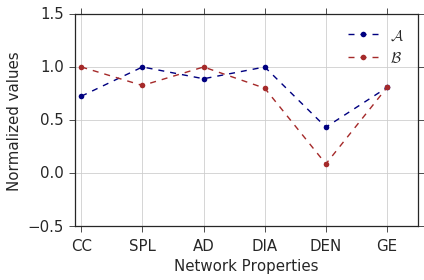}\\
 (a)&(b)
 \end{tabular}
 \caption{The figure showing (a) spectra of Les Miserables network and that of the subset with $q = 38$, (b) normalized network properties of original network and that of subset network. We can see that the SPL is smaller and CC and GE are larger in subset network than in the original network.}
\label{fig:cs7}
 \end{figure*}
 
 \begin{figure*}[htbp]
\centering
 \includegraphics[height=6cm, trim={40 60 40 60},clip]{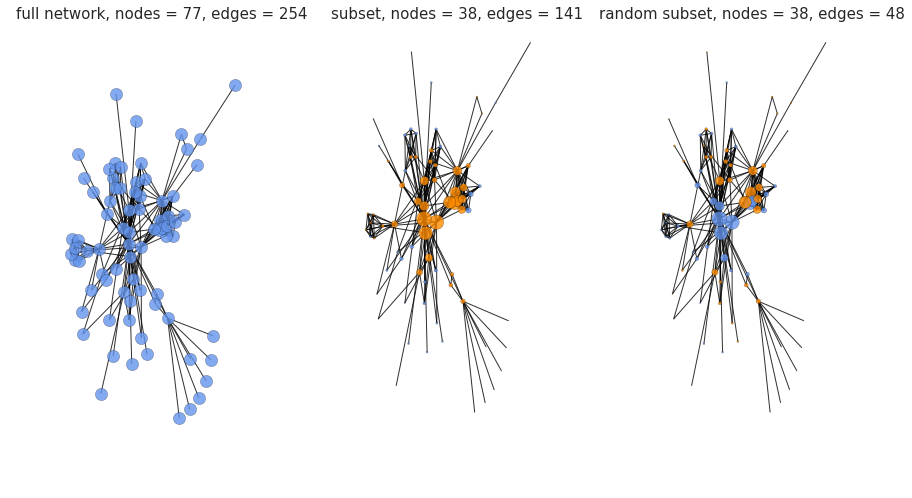}
 \caption{The figure showing the Les Miserable network (first panel), with subset embedded in the network in orange colour (second panel) and a random subset embedded in network in orange colour (third panel). All the nodes in second and third panel are scaled in size according to PSV components values.}
\label{fig:cs8}
 \end{figure*}

\begin{table*}
\label{tab:tab1}
\begin{center}
    \begin{tabular}{| p{2.7cm} || p{2.3cm} | p{2cm} | p{1cm} | p{2cm} | p{2cm} | p{1cm} | p{3cm} |}
    \hline
    Type &Networks &  (V, E) of network  & q & (V, E) of SS network & $ \frac{\parallel \mathcal{A} \parallel- \parallel \mathcal{A}_1\parallel}{\parallel \mathcal{A} \parallel} $ & CS & IPR and SS contribution to IPR\\
    \hline
    \multirow{3}{*}{Weighed, real}& US Air & (332, 2126) & 166 &                                       (890,1635) & 0.011 & 0.99                                      &(0.0255, 0.0255)\\
                       &Les Miserables & (77, 254) & 38 & (38, 141) & 0.017 & 0.99 & (0.1176, 0.1176)\\
                       &Train Bombing & (64, 243) & 32 & (32, 123) & 0.100 & 0.99 & (0.0906, 0.0886)\\
                       \hline
    \multirow{6}{*}{Unweighted, real}&Karate & (34, 78) & 20 & (20,                                      47)& 0.108 & 0.97 &                                         (0.073, 0.069) \\
                        &Cat Brain & (65, 730) & 32 & (32, 247) & 0.236 & 0.99 &(0.024, 0.017)\\
                        &Drosophila & (1781, 9016) & 890 & (890, 7026)  & 0.0572 & 0.99 & (0.048, 0.048)\\
                        &Power Grid & (4941, 6594) &  2470 & (2470, 2863) & 0.1668 & 0.985 & (0.0408, 0.0344)\\
                        &Jazz Musicians& (198, 2742) & 96 & (96, 947) & 0.2317 & 0.98 & (0.014, 0.010)\\
                        &Friendship & (1858, 12534) & 929 & (929, 7618) & 0.114 & 0.99 & (0.010, 0.009)\\
                        \hline
    \multirow{4}{*}{Unweighted, model}&Barabasi Albert & (1000, 2991) & 500 & (500,                                   1418) & 0.1503 & 0.99 &                                      (0.026, 0.026) \\
                        &Erdos Renyi & (1000, 7558) & 500 & (500, 2630) & 0.232 & 0.98 & (0.0013, 0.0011) \\
                        &Power Law & (1000, 1360) & 500 & (500, 947) & 0.079 & 0.99 & (0.0595,0.0594) \\
                        &LFR & (1000, 2929) & 500 & (500, 1537)& 0.127 & 0.88 & (0.144, 0.144) \\
   
    \hline
    \end{tabular}
\end{center}
\caption{A table of SS results on model networks and weighted and unweighted real networks examples. The real networks were downloaded from KONECT \cite{kunegis2013konect} and model network types were generated using python module Networkx \cite{hagberg2008exploring}. (V, E) represents the vertices and the edges in the networks. CS represents cosine similarity between PSVs of main network adjacency and the subset.}.
\end{table*}
 
\subsection{Barabasi-Albert model Network}
We present similar results for a Barabasi-Albert network constructed using Networkx Python library \cite{hagberg2008exploring} with total $1000$ nodes and value $m$ representing number of pre-existing nodes a new node makes connection to as $3$. For SS, $q = 500$ was chosen. The matrix norm of original and subset matrices are $77.34$ and $65.71$. It is seen that even when only $50 \%$ of columns were selected the CS between PSVs is $0.99$. \\

The schematic of the embedded subset (nodes in orange color) in FIG.~\ref{fig:cs6}, shows that subset networks comprise of nodes that are well connected to that of rest of the network and have high eigenvector centralities (node sizes are scaled with eigenvector centrality). The edges retained in the actual subset and the random subset are $1418$ and $763$ respectively. The IPR values of the original network were $0.0266$; the contribution of the nodes exclusively in the subset and random subset are, $0.0265$ and $0.0164$ respectively. This indicates that the nodes captured in the subset contribute to IPR maximally. This is a significant result as it is indicative of the presence of unimportant nodes in the original network. \\

The comparison of network properties shows that the subset network has higher CC, lower SPL, and higher GE than the original networks (FIG.~\ref{fig:cs4}(b)). The CC, SPL, GE of the main network were $0.0263, 3.51, 0.301$ and that of subset network was $0.030, 3.48, 0.308$. This observation demonstrates that subset represents an effective network subgraph (the nodes and edges) that plays an essential part in information flow across the network.\\

\subsection{Les Miserables Network}
Les Miserables is a weighted network of 77 nodes, and 254 edges where nodes are the characters in novel Les Miserables by Victor Hugo and weighted edges define number their co-occurrences in a chapter. We find that the selected subset for the weighted networks approximates the spectra of the original network to the best degree (see FIG.~\ref{fig:cs7}(a) and table). For this network, the Frobenius norms of the main network matrix and the subset (with $q = 38$) were $109.23, 107.34$. The CS of PSV of main network adjacency and the subset was $0.99$. The subset captures all the nodes with higher components values of PSV as opposed to a random subset (FIG.~\ref{fig:cs8}). The IPR values of the original network were $0.11769$; the contribution of the nodes exclusively in the subset and random subset are, $0.11767$ and $0.01401$ respectively.  Also, the network properties governing efficiency of information flow were significantly enhanced in subset network as can be seen from the figure~\ref{fig:cs7}(b).\\

We obtained these statistics for other real world and model network types as well (refer TABLE I.). We find that the loss in matrix energy is minimum and CS of PSVs of $\mathcal{A}$ and  $\mathcal{A}_1$ is maximum for the case of weighted networks as compared to unweighted real and model networks.

\section{Conclusions}
In this work, we have presented the outcomes of the application of the subset selection algorithm on the complex network adjacency matrices. The subset selects the columns are that are most linearly independent, from a matrix. Hence, the subset comprises of vectors that can span the whole column space of the matrix. We found that the subset selected from main network adjacency matrix comprises of important network substructure. The observation confirms the importance of the selected subset that nodes in subset were the ones that had the highest eigenvector centrality in the main network. We also observe that the subset network is the connected one as compared to any other random subset selected (of the same size) from the main network. This was observed when the subset selection extracted just 50$\%$ of the nodes and their interconnections. Also, in almost all of the subset networks, the information flow statistics were improved than the main network.
For example, the clustering coefficient was higher; the average shortest path length was lower and global efficiency was higher in the subset than the main network. This observation is suggestive of the two things (a)the subset extracts the most functional network sub-part and (b) original network is more robust to failure of nodes that are there in the redundant set. However, the second point can be only be stated with confidence if the value $q$ was chosen correctly. The choice of $q$, such that the spectra are correctly reproduced in the subset is strongly influenced by network sparsity; more sparse the network is maximum number of column vectors may be there which have all zeros, whereas if the network is dense, most of the columns will be non-identical and hence greater $q$ is required for subset.\\
We found that the subset reproduces the spectral properties of the main network the best for the case of weighted networks. This is intuitively obvious as the all column vectors having combinations of $0$s and $1$s tend to serve as orthogonal vectors if they are non-repeating.  For a non-binary or weighted matrix, the linearly independent columns can serve as a basis for a large number of redundant columns which can be written as a linear combination of these set of columns. It is expected for a weighted adjacency matrix, the number of linearly independent columns will be very less in comparison to the number of columns of an adjacency matrix. Therefore the size of the reduced adjacency matrix will be small, and hence it is easy to handle such small matrices, and it can be useful to analyze the spectral properties of the networks without analyzing the original large adjacency matrix.\\
We also demonstrated the subset selection on a few real-world networks and found exciting results. Our proposal is also a good approach to finding important or influential network nodes. We hope that this approach allows researchers to analyze even larger networks with millions of nodes such as the internet and the World Wide Web and find the most functional network structure and hence the nodes. We look forward to such applications.

\bibliography{References}

\end{document}